\documentclass[aip,jcp,twocolumn,citeautoscript,floatfix,reprint]{revtex4-1}

\usepackage{graphicx}
\usepackage{amssymb}
\usepackage{amsbsy}
\usepackage{multirow}
\usepackage{mathrsfs}
\usepackage{amsmath}
\usepackage[T1]{fontenc}
\DeclareMathAlphabet{\mathscrbf}{OMS}{mdugm}{b}{n}
\usepackage{bm}
\usepackage{color}
\usepackage{epstopdf}
\usepackage[normalem]{ulem}
\usepackage{verbatim}
\usepackage[table]{xcolor}

\usepackage{float}
\restylefloat{table}
\usepackage{placeins}
\usepackage{booktabs}
\usepackage{cleveref}

\newcommand{\beq}{\begin{equation}}
\newcommand{\eeq}{\end{equation}}

\newcommand{\Trm}{{\bf T}}
\newcommand{\Crm}{{\bf C}}

\newcommand{\FoT}{\mathscr{F}({\bf T})}
\newcommand{\F}{\mathscr{F}}

\newcommand{\At}{{\rm Ala}_{3}}
\newcommand{\Aq}{{\rm Ala}_{4}}
\newcommand{\Rof}{R_{\operatorname{1-4}}}

\begin{document}
\title{Consistent Interpretation of Molecular Simulation Kinetics Using Markov State Models Biased with External Information}
%Improved Kinetics from Molecular Simulation Trajectories using Biased Markov State Models
%Improved Kinetics of Molecular Simulation Trajectories using Biased Markov State Models
% Improving Molecular Simulation Kinetics with Biased Markov State Models
% Reproducing Kinetic Observables with Biased Markov State Models
\author{Joseph F. Rudzinski}
\email{rudzinski@mpip-mainz.mpg.de}
\author{Kurt Kremer}
\author{Tristan Bereau}
\affiliation{Max Planck Institute for Polymer Research,  55128 Mainz, Germany}
%\date{\today} % Activate to display a given date or no date

\begin{abstract}
Molecular simulations can provide microscopic insight into the physical and chemical driving forces of complex molecular processes.  
%Despite continued advancement of simulation methodology, model errors may lead to inconsistencies between simulated and experimentally-measured observables.  
Despite continued advancement of simulation methodology, model errors may lead to inconsistencies 
between simulated and reference (e.g., from experiments or higher-level simulations) observables.
To bound the microscopic information generated by computer simulations within reference measurements, we propose a method that reweights the microscopic transitions of the system to improve consistency with a set of coarse kinetic observables.  
The method employs the well-developed Markov state modeling framework to efficiently link microscopic dynamics with long-timescale constraints, thereby consistently addressing a wide range of timescales.
To emphasize the robustness of the method, we consider two distinct coarse-grained models with significant kinetic inconsistencies.  
When applied to the simulated conformational dynamics of small peptides, the reweighting procedure systematically improves the timescale separation of the slowest processes.  
Additionally, constraining the forward and backward rates between metastable states leads to slight improvement of their relative stabilities and, thus, refined equilibrium properties of the resulting model.
Finally, we find that difficulties in simultaneously describing both the simulated data and the provided constraints can help identify specific limitations of the underlying simulation approach.
\end{abstract}

%\clearpage 
\maketitle
%\thispagestyle{empty}


%\setcounter{page}{1}

Despite acknowledged limitations in current all-atom (AA) force fields to describe complex molecular systems (e.g., proteins\cite{Best:2008,Cino:2012}), the confidence associated with atomically-detailed molecular dynamics  simulations continues to increase.\cite{Beauchamp:2012} 
This can be attributed to improvements in simulation models\cite{Best:2009, Lindorff-Larsen:2010, Best:2012,Jiang:2013} and methodologies,\cite{Shaw:2010fk, Zuckerman:2011, Markwick:2011, Fujisaki:2015, Morriss-Andrews:2015} as well as continued experimental validation.\cite{Lange:2010, Lindorff-Larsen:2012} 
The latter has been facilitated by both increased resolution of experiments\cite{Nettels:2007, Chung:2012, Oikawa:2013, Otosu:2015} and improved tools\cite{Noe:2011, Keller:2012, Lindner:2013, Yi:2013} for comparing simulated and measured data.
%Beyond ongoing improvements of simulation models, existing simulation trajectories can be altered to improve their agreement with known, external data.\cite{Groth:1999,Rozycki:2011}
Beyond ongoing improvements of simulation models, the microscopic insight extracted from existing simulations can be refined by altering the trajectories to improve their agreement with external data.\cite{Groth:1999,Rozycki:2011}
For example, Beauchamp \emph{et al.}\cite{Beauchamp:2014} recently proposed a method to reweight the ensemble of peptide configurations generated from a molecular dynamics simulation to be consistent with experimental chemical shift and $^3J$ measurements, leading to a systematic improvement of secondary-structure propensities. 

Expanding upon this idea, the present work proposes a method to improve the kinetic properties determined from a simulation, given a set of reference observables.  
%This approach may be useful for, e.g., addressing the kinetic discrepancies of AA force fields identified in recent investigations of extensively-sampled peptide trajectories.
%This approach may be useful for, e.g., characterizing the relationship between dynamics generated by AA and coarse-grained models. 
%This relationship is generally not well understood, limiting the applicability of a large number of CG models to static equilibrium properties.\cite{Peter:2009hr,Noid:2013uq}
We seek to relate the microscopic transitions of the system with much coarser observables, effectively linking a wide range of timescales.  
Practically, this link is provided by Markov state models (MSMs), which describe the long-time dynamics of a system with a memoryless evolution of microstate transitions.
The methodology for constructing MSMs directly from simulations has been extensively developed\cite{Chodera:2006, Noe:2009vn, Prinz:2011, Chodera:2014, Nuske:2014, Bowman:2014, Wu:2015, Schutte:2015, Vitalini:2015b} and MSMs are routinely employed to elucidate complex simulated processes, e.g., protein folding\cite{Chodera:2007,Noe:2007,Prinz:2011b,Chodera:2011,Lane:2011,Bowman:2011uq} and protein-ligand binding.\cite{Beauchamp:2011,Metzner:2011,Buch:2011,Bowman:2012,Plattner:2015}
Additionally, recent work\cite{Vitalini:2015} has applied MSMs to identify various discrepancies in the dynamical properties generated by different AA force fields.
%By incorporating both simulation data as well as experimental kinetic constraints, the proposed method may be useful for investigating the relevance and source of such discrepancies.
By incorporating simulation data as well as experimental kinetic constraints, the proposed method may provide insight into the relevance and source of such discrepancies.
Moreover, the approach may be particularly useful to characterize the relationship between dynamics generated by AA and coarse-grained (CG) models. 
This relationship is generally not well understood, limiting the applicability of a large number of CG models to static equilibrium properties.\cite{Peter:2009hr,Noid:2013uq}

An MSM is fully characterized by a transition probability matrix, $\Trm(\tau)$, whose elements, $T_{ij}$, describe the probability of jumping from microstate $i$ to $j$ within a ``lag time'' $\tau$.  
The number of observed jumps between each pair of microstates during a simulation determines the count matrix, $\Crm^{\rm{obs}}(\tau)$.  
An MSM that accurately describes the long-time simulation dynamics may be constructed by maximizing the log-likelihood function $Q(\Trm) = \ln p(\Trm \mid \Crm^{\rm{obs}})$, where the probability of the model given the simulation data is\cite{Prinz:2011b}
\begin{equation} 
  p(\Trm \mid \Crm^{\rm{obs}}) \propto p(\Crm^{\rm{obs}} \mid \Trm) = \prod_{ij} T_{ij}^{C^{\rm{obs}}_{ij}}.
\label{eq-1}
\end{equation} 
The proportionality follows from Bayes' theorem, while the rightmost expression is implied by Markovian dynamics.
The resulting maximum likelihood estimate (mle), $\Trm^{\rm mle}$, represents the MSM most likely to generate $\Crm^{\rm{obs}}$.
Physical constraints, e.g., detailed balance, are typically incorporated\cite{Noe:2008} into the optimization of $\Trm^{\rm mle}$ in order to overcome finite sampling errors of the simulation.  
Moreover, convex optimization routines\cite{Prinz:2011b} can efficiently determine $\Trm^{\rm mle}$ for MSMs with hundreds of microstates.\cite{Plattner:2015}  

In the present work, we consider the mle problem (Eq.~1), while incorporating kinetic constraints, $\FoT = 0$, between macrostates (i.e., collections of microstates).   
Unfortunately, these constraints are, in general, nonlinear functions of the elements $T_{ij}$, preventing the straightforward application of Lagrange multipliers or the use of convex optimization routines.\cite{Optimization:1995}  
Furthermore, a simulation model may prove incompatible with $\FoT$, such that strict enforcement would destroy much of the microscopic information provided by the trajectory.  
As such, we seek to maximize the agreement with the constraints while minimally biasing the original MSM (i.e., $\Trm^{\rm mle}$).  
We achieve this balance with Metropolis Monte Carlo sampling of transition probability matrices according to
\begin{equation} 
E_{\rm tot}(\Trm \mid \lambda) = \lambda E_Q(\Trm) +
(1-\lambda) E_{\F}(\Trm),
\label{eq-2}
\end{equation} 
where the energies $E_Q(\Trm)$ and $E_{\F}(\Trm)$ are shifted and rescaled quantities with respect to ${\rm Q}(\Trm)$ and $\FoT$, respectively.\cite{SI}
These quantities are defined such that $E_Q(\Trm^{\rm mle}) = 0$, $E_{\F}(\Trm) = 0$ when $\FoT = 0$, and max($E_Q(\Trm)$), max($E_{\F}(\Trm)$) $\approx 1$ for the relevant range of sampled matrices.  
The control parameter, $\lambda$, balances the contribution of the two quantities.  
In practice, we monitor the two energy terms while tuning $\lambda$ in order to determine the optimal ``biased'' MSM.
We note that when $\lambda = 1$ the scheme provides the uncertainty of $\Trm^{\rm mle}$.\cite{Noe:2008,Metzner:2009,Metzner:2010}  
While we include additional technical details in the Supporting Information,\cite{SI} we leave a detailed assessment to a subsequent publication.  
All MSM calculations employed an inhouse extension of the {\sc pyEmma} package.\cite{PyEmma,Senne:2012}

We apply the proposed method to the conformational dynamics of two small peptides.  
For each system, we consider simulations of both an AA and a CG model.
To illustrate the robustness of the method, we consider a highly specific bottom-up model for one system while using a more transferable top-down model for the other.
CG models, which lump several atoms into a single CG site, often display faster dynamics than a corresponding AA model due to reduced molecular friction between sites.  
%Unfortunately, the relationship between CG and AA dynamics is generally not well understood,\cite{Peter:2009hr,Noid:2013uq} limiting the applicability of a large number of CG models to static equilibrium properties.
As a consequence, these models are excellent candidates to test the present methodology.

For each system, we compare three distinct MSMs constructed from the simulations: 
(1) the ``ref'' model: an unbiased MSM constructed using AA data; 
(2) the ``UMSM'': an unbiased MSM constructed using CG data; 
and (3) the ``BMSM'': a biased MSM constructed via Monte Carlo sampling
%according to Eq.~\ref{eq-2}, which incorporates both the CG simulation trajectory as well as external constraints.  
according to Eq.~2, which incorporates CG simulation data as well as external constraints.  
The chosen constraints involve mean first passage times (MFPTs), $\{m_K\}$, between metastable states determined from the ref model.
The MFPTs are calculated directly from $\Trm$ by solving a set of linear equations.\cite{PyEmma,Senne:2012}
Here, the ref model allows a detailed assessment of the properties of the UMSM and BMSM.  
In general, however, a reference MSM is unnecessary, since the procedure requires only coarse (e.g., macrostate-level) information.

As expected, the CG models displayed significantly faster dynamics than the underlying AA model, allowing enhanced sampling at reduced computational expense.  
Ideally, a CG model should retain enhanced dynamics while consistently or predictably speeding up all relevant kinetic processes.   
To this end, we consider only ratios of MFPTs: $\tilde{m}_K \equiv m_{K} / m_{\rm L}$, where $K$ denotes a (directional) transition between two metastable states and L denotes the particular transition corresponding to the longest MFPT of the ref model.
We set the constraint as the root sum square of relative errors of MFPT ratios, 
$\F = \sqrt[]{\sum_{K} \left( \tilde{m}_{K} - \tilde{m}^{\rm ref}_{K} \right )^2 / \left(\tilde{m}^{\rm ref}_{K} \right )^2} $.
Thus, we aim to recover the dynamics of the system up to a \emph{homogeneous} speedup factor, which requires $\F = 0$.  
We circumvent a calibration of the AA and CG timescales by only comparing the eigenvalues of the MSMs, $\{ \lambda_j \}$, which are linked to the timescales of particular processes $\{ j \}$ by $t_j = -\tau/\ln \lambda_j$.\cite{Bowman:2014}
Due to the ambiguity of the CG dynamics, we report CG timescales in reduced units, $\mathcal{T}^{\rm CG}$, specific to the model.

% Figure 0
\begin{figure}[htb]
\includegraphics[width=\linewidth]{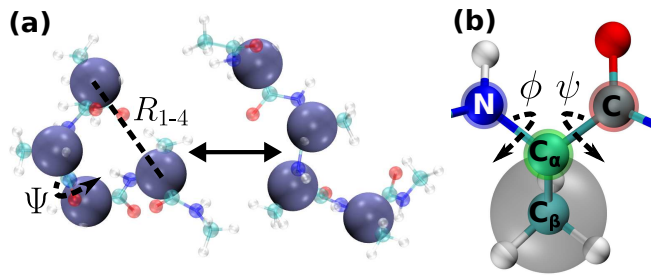}
\caption{Representation and relevant degrees of freedom of the CG models for
  (a) $\Aq$ and (b) $\At$.  Rendered with VMD.\cite{Humphrey:1996}
  }
\label{Fig-0}
\end{figure}

% ALA-4 
We first considered a tetra-peptide of alanine residues ($\Aq$).  The AA
simulation employed the OPLS-AA\cite{Jorgensen:1996} and
SPC/E\cite{Berendsen:1987} force fields to model an explicitly solvated,
capped $\Aq$ peptide.
The CG simulation employed a structure-based force
field,\cite{Rudzinski:2014b} which represents each amino acid with a single CG
site placed at the $\alpha$-carbon position, to model an implicitly solvated
$\Aq$ peptide.  
This model qualitatively reproduces the free-energy surface (FES) along the
dihedral angle, $\Psi$, defined between the four $\alpha$-carbons of the
peptide backbone and the end-to-end distance, $\Rof$, between the first and
last $\alpha$-carbons (Fig.~\ref{Fig-0}a).  The simulation and
parameterization details were previously published.\cite{Rudzinski:2014b} Both
AA and CG trajectories were discretized on a uniform
grid along $\Psi$ and $\Rof$.  MSMs were constructed 
with lag times of $\tau =$ 250~ns and 1.25~$\mathcal{T}^{\rm S}$ for the AA and CG
models, respectively, where $\mathcal{T}^{\rm S}$ denotes the time unit for the structure-based model. 

% Figure 3
\begin{figure}[htb]
\includegraphics[width=\linewidth]{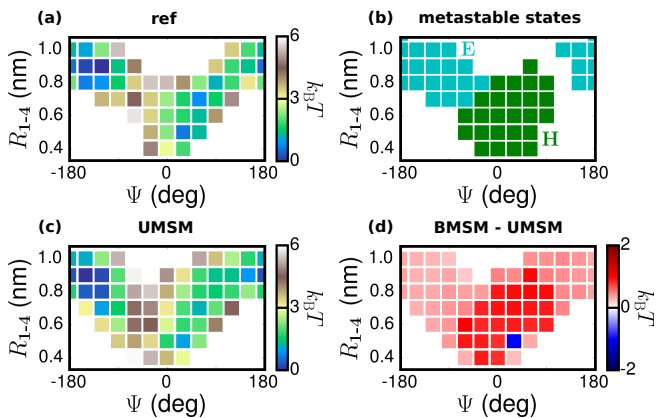}
\caption{%\cite{Hunter:2007} 
	$\Aq$. Free-energy surfaces of $\Aq$ along $\Psi$ and $\Rof$ determined from
  the (a) ref model, (c) UMSM, and (d) difference between the BMSM and UMSM.  (b)
  Helical (H, green) and extended (E, cyan) metastable states. }
\label{Fig-3}
\end{figure}

\begin{table}[htb] \normalsize
\caption{ Free-energy differences (in units of $k_{\rm B} T$) with respect to
the most stable metastable state for (a) $\Aq$ and (b) $\At$.    } \centering
\vspace{5pt}
\begin{tabular}{|c||c||c|c|} \multicolumn{1}{l}{} & \multicolumn{1}{c||}{(a)
$\Aq$} & \multicolumn{2}{c}{(b) $\At$} \\ \cline{2-4}
\multicolumn{1}{l|}{} & $\rm{H\operatorname{-}E}$ & $\alpha \operatorname{-}
\beta$ & $\alpha_{\rm L} \operatorname{-} \beta$ \\ \cline{2-4} \hline ref & $0.51$
& $0.42$ & $4.18$ \\ \hline UMSM & $0.96$ & $0.95$ & $5.75$ \\ \hline BMSM &
$0.88$ & $0.75$ & $5.73$ \\ \hline
\end{tabular}
\label{Tab-1}
\end{table}

Fig.~\ref{Fig-3} compares the FESs along $\Psi$ and $\Rof$ determined from the ref model (panel a) and the UMSM (panel c).  
Additionally, panel b presents two metastable states, determined from the ref model via the PCCA+ algorithm,\cite{Deuflhard:2005} corresponding to helical (H, green) and extended (E, cyan) structures.  
The AA model connects these two regions through intermediates with $\Psi \approx$ 130~deg and $\Rof \approx$ 0.9~nm.
While it is possible to describe the transition in more detail, we only consider these two states to focus on the description of the slowest process.
Table~\ref{Tab-1}a presents the free-energy difference between the H and E states for each model.  
Despite the apparent structural agreement of the FESs, the free-energy difference between the metastable states is significant.

\begin{figure}[htb]
\includegraphics[width=\linewidth]{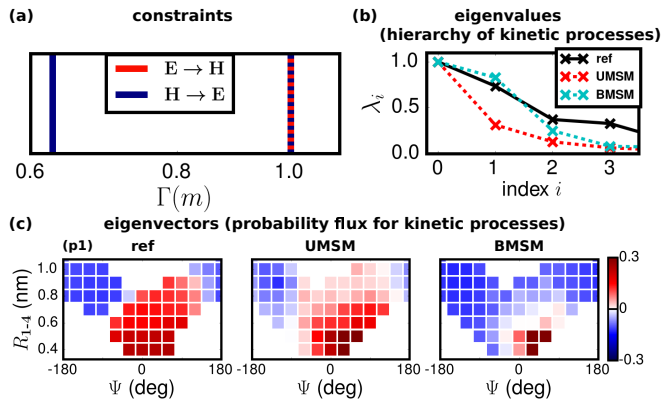}
\caption{%\cite{Hunter:2007} 
	$\Aq$. (a) Relative fractional speedup, $\Gamma$ (defined in text), of each MFPT for the UMSM
  (solid lines) and BMSM (dashed lines); (b) 4 largest eigenvalues; and (c)
  Eigenvector characterizing the slowest dynamical process ($\lambda_1$) of
  each model.
  The intensity plots describe a flux of probabilities between microstates with positive and negative values, weighted by the magnitude of the individual components.
  }
\label{Fig-4}
\end{figure}

Fig.~\ref{Fig-4} demonstrates that there are also significant discrepancies in
the kinetic properties of the UMSM.  
We first probe the accuracy of the ratios of MFPTs between the metastable states, which will be employed as constraints in the construction of the BMSM.  
This agreement is assessed by calculating the relative fractional speedup, $\Gamma$, of each MFPT: 
$\Gamma( m_{K} ) \equiv \tilde{m}_{K} / \tilde{m}^{\rm ref}_{K}.$
Deviations from $\Gamma = 1$ indicate discrepancies in the MFPTs, beyond a homogeneous speedup factor.  
The solid blue line in Fig.~\ref{Fig-4}a demonstrates that the ${\rm H} \rightarrow {\rm E}$ transition is too fast compared to the reverse process in the UMSM.

The kinetic properties of the UMSM are further characterized from the eigenvalues and eigenvectors of the transition probability matrix.  
Fig.~\ref{Fig-4}b presents the largest four eigenvalues of each model.  
For any MSM of a system at equilibrium, the largest eigenvalue is $\lambda_0 = 1$ and its eigenvector coincides with the equilibrium probability distribution.\cite{Bowman:2014} 
The remaining eigenvectors describe the slowest processes of the system, sorted by their eigenvalue.
Fig.~\ref{Fig-4}c presents the eigenvector of $\lambda_1$ for each model, 
which describes a flux of the probability distribution between microstates with positive and negative values, weighted by the individual eigenvector component of each microstate. 
The UMSM properly describes the transition between the two metastable states, likely due to the careful parametrization of the CG model.\cite{Rudzinski:2014b} 
On the other hand, Fig.~\ref{Fig-4}b demonstrates that the UMSM does not reproduce the implied separation of timescales between indices 1 and 2 of the ref model.

% Figure 4

Starting with the UMSM, the constraint function, $\F$, built from the MFPT ratios between states ${\rm H}$ and ${\rm E}$, was applied to sample transition matrices according to Eq.~2.  
From the ensemble of MSMs approximately fulfilling the constraint, an optimal BMSM, sampled with $\lambda = 0.875$, was chosen to balance $E_Q(\Trm)$ and $E_{\F}(\Trm)$ (see Fig.~S7\cite{SI}).
The blue dashed line in Fig.~\ref{Fig-4}a demonstrates that the resulting BMSM nearly quantitatively reproduces the given constraint, i.e., $\Gamma(m_{\rm E \rightarrow H}) \approx 1$.  The BMSM yields a larger separation of timescales (Fig.~\ref{Fig-4}b), in much better agreement with the ref model.  
However, the description of the slowest process (Fig.~\ref{Fig-4}c) is somewhat degraded, with increased probability flux in a narrow region of ${\rm H}$.
At the same time, the most probable microstate in this region is significantly stabilized (Fig.~\ref{Fig-3}d). 

The non-uniform distribution of probability flux characterizing the H to E transition is already apparent in the UMSM's description of process 1 (Fig.~\ref{Fig-4}c).
Moreover, further analysis (Fig.~S7 and S8\cite{SI}) indicates that this feature is not an artifact of sampling but, rather, emerges systematically from the combined application of the CG simulation and reference data to construct the BMSM.
In Eq.~2, $E_Q$ ensures the essential dynamical features of the underlying simulation are minimally perturbed.
The inclination of the BMSM to reproduce the MFPTs and timescale separation by exacerbating the concentrated probability flux of the helical region implicates this non-uniform flux as an essential component of the underlying simulated processes.

Interestingly, previous work demonstrated that the CG interaction potentials of this model stabilize helical transitions through strong, non-cooperative interactions in order to compensate for the presence of conformations sterically forbidden in the AA model.\cite{Rudzinski:2014b}
This feature appears to give rise to a transition from helix to extended structures that is inherently more localized within the helical state, leading to the heterogeneous flux profiles observed in the CG MSMs.
Although further investigation is required to clarify the precise connection between these features, this analysis strongly implicates the methodology as a useful tool for identifying inherent limitations in the underlying simulation model.

% ALA-3 
As a second example, we also considered a tri-alanine peptide ($\At$).  The AA
simulation employed the same model and implementation as described above for
$\Aq$.  The CG simulation employed the top-down PLUM force
field,\cite{Bereau:2009} which represents each heavy atom of the peptide
backbone as well as each side chain with a CG site, to model an implicitly
solvated $\At$ peptide.  Both trajectories were discretized on a uniform 
grid along the $\phi$ and $\psi$ dihedral angles of the center residue
(Fig.~\ref{Fig-0}b).  MSMs were then constructed  
with lag times of $\tau =$ 40~ns and 1.5~$\mathcal{T}^{\rm P}$ for the AA and CG
models, respectively, where $\mathcal{T}^{\rm P}$ denotes the time unit for the PLUM model. 

% Figure 1
\begin{figure}[htb]
\includegraphics[width=\linewidth]{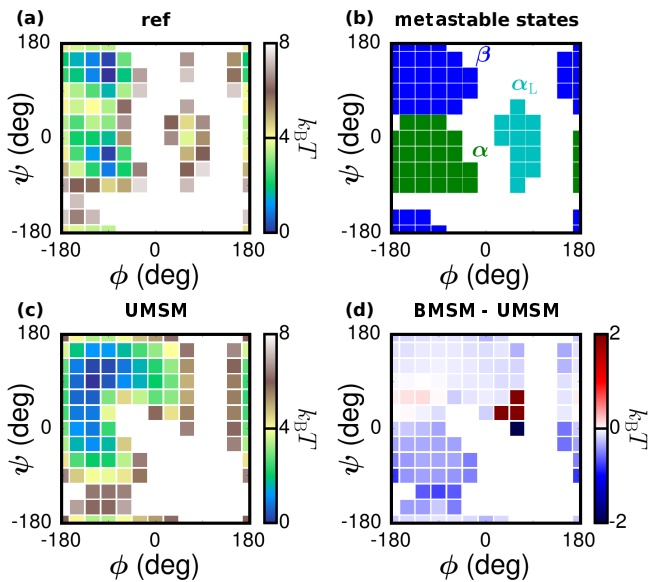}
\caption{%\cite{Hunter:2007} 
	$\At$. Free-energy surfaces of $\At$ along the $\phi$ and $\psi$ dihedral
  angles determined from the (a) ref model, (c) UMSM, and (d) difference
  between the BMSM and UMSM.  Panel (b) presents the definition of the three
  metastable states: corresponding to alpha-helical ($\alpha$, green),
  beta-sheet ($\beta$, blue), and left-handed-helical ($\alpha_{\rm L}$, cyan)
  regions.  }
\label{Fig-1}
\end{figure}

\begin{figure}[htb]
\includegraphics[width=\linewidth]{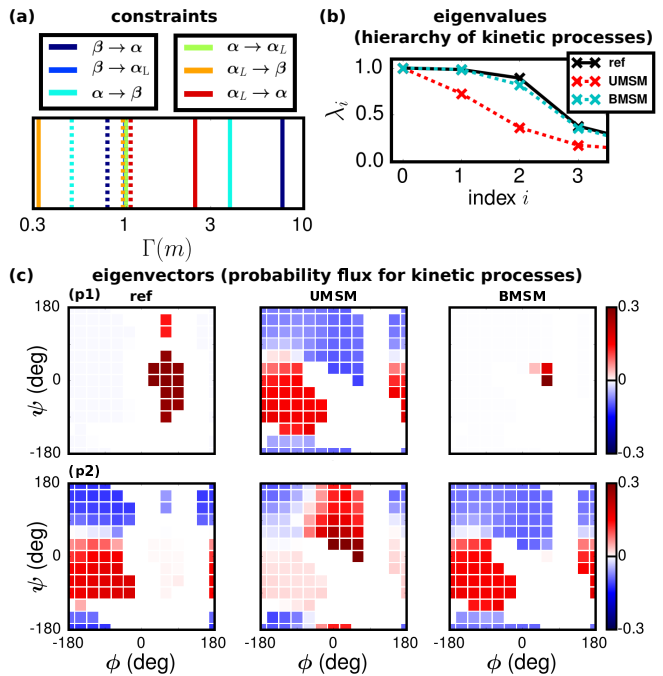}
\caption{ $\At$. (a) Relative fractional speedup, $\Gamma$ (defined in text), of each MFPT for the UMSM
  (solid lines) and BMSM (dashed lines); (b) 4 largest eigenvalues; and (c)
  Eigenvectors characterizing the two slowest dynamical processes, $\lambda_1$
  (p1) and $\lambda_2$ (p2) of each model. 
  (Negative values are too small to be noticed in p1 ref and
BMSM).
  }
\label{Fig-2}
\end{figure}

Fig.~\ref{Fig-1} compares FESs along $\phi$ and $\psi$ determined from the ref model (panel a) and the UMSM (panel c).  
Additionally, panel b presents three metastable states, determined from the ref model via the PCCA+ algorithm,\cite{Deuflhard:2005} corresponding to alpha-helical ($\alpha$, green), beta-sheet ($\beta$, blue), and left-handed-helical ($\alpha_{\rm L}$, cyan) structures.  
Panel c shows that the CG model samples the metastable states with incorrect propensities (quantified in Table~\ref{Tab-1}b).  
In terms of kinetics, the solid lines in Fig.~\ref{Fig-2}a indicate that the timescales of transition between the metastable states differ qualitatively from the ref model, beyond a homogeneous speedup factor.  
In this case, $\Gamma$ is determined relative to the $\alpha \rightarrow \alpha_{\rm L}$ transition.
%(the longest process in the ref model).  
Fig.~\ref{Fig-2}b demonstrates that the UMSM does not reproduce the implied separation of timescales between indices 2 and 3.  
%Worse still, the eigenvectors of $\lambda_1$ and $\lambda_2$ are qualitatively different from the ref model, indicating a distinct hierarchy of kinetic processes (Fig.~\ref{Fig-2}c).
Worse still, because $\lambda_1$ and $\lambda_2$ are nearly degenerate, the hierarchy of kinetic processes cannot be reliably determined.
In this case, the order of these processes is qualitatively different from the ref model (Fig.~\ref{Fig-2}c).
While the two slowest processes of the ref model (column 1) correspond to transitions involving $\alpha_{\rm L}$ and between $\alpha$ and $\beta$, the processes of the UMSM (column 2) are in reverse order and significantly skewed.

% Figure 2

Similar to $\Aq$, the constraint $\F$, characterizing the error in the MFPT ratios between metastable states, was applied to sample MSMs according to Eq.~2.
An optimal BMSM was identified as described above for $\Aq$, sampled with $\lambda = 0.9$ (see Fig.~S10\cite{SI}).  
In this case, the constraints could not be perfectly fulfilled without significantly deteriorating agreement with the simulation data.
The dashed lines in Fig.~\ref{Fig-2}a quantify this discrepancy.
Fig.~\ref{Fig-1}d demonstrates the BMSM's pronounced, but localized, adjustments of microstate stabilities in the $\alpha_{\rm L}$ region, although the overall stability remains largely unchanged (Table~\ref{Tab-1}b).  
%The BMSM also slightly destabilizes the region at the interface between the $\alpha$ and $\beta$ metastable states, while providing an overall stabilization of both regions, resulting in a significantly improved free-energy difference (Table~\ref{Tab-1}b).
The BMSM also slightly destabilizes the interface region between the $\alpha$ and $\beta$ metastable states, while providing an overall stabilization of both regions, resulting in a significantly improved free-energy difference (Table~\ref{Tab-1}b).
These adjustments result in relative fractional speedups much closer to 1 (Fig.~\ref{Fig-2}a, dashed lines) as well as excellent agreement in the separation of timescales (Fig.~\ref{Fig-2}b).  
%Moreover, not only are the eigenvectors associated with the first two processes more accurately described, but the hierarchy is also restored (Fig.~\ref{Fig-2}c).
Moreover, not only do the first two eigenvectors more accurately describe the underlying processes, but the hierarchy is also restored (Fig.~\ref{Fig-2}c).

% Summary and Conclusions

This work outlines a simple method to determine an MSM that combines information from a computer simulation with a set of kinetic constraints. 
Importantly, the scheme does not severely restrict the form of the constraint, allowing experimental measurements to inform the construction of the model.  
The proposed framework also allows simple and transparent flexibility in the enforcement of the constraints.  
Indeed, we find that the optimal model may not perfectly reproduce the given constraints.  
For the conformational dynamics of two small peptides, the BMSM improves the description of kinetics, both in terms of the constrained MFPTs and the implied timescales associated with the slowest processes, while refining slightly but systematically the equilibrium distribution of the metastable states.

In the context of CG models, the method provides a systematic framework to interpret kinetic properties in a meaningful and consistent way.
%We envision to ``scale up'' the reweighting of microscopic transitions beyond the system used for the BMSM calculation, and study the associated transferability.
We plan to investigate the transferability of micro-trajectory reweighting beyond the system used in the original calculation.
%Interestingly, we also find that the BMSM may exacerbate artifacts of the underlying model, implicating the method as a potential tool to identify specific limitations of simulation models and possibly refine them.
Interestingly, we find that the BMSM may exacerbate artifacts of the underlying model, implicating the method as a potential tool for the refinement of molecular force fields, in the case that an underlying model is available.
Finally, we expect the method will also be useful for investigations comparing high-resolution AA simulations and experimental measurements of complex biomolecular processes, e.g., protein folding.

\section*{Acknowledgments}

The authors thank Will Noid for the use of the $\Aq$ simulation trajectories.
We thank Denis Andrienko, Cristina Greco, and Marc Radu for critical reading of the manuscript and Benjamin Trendelkamp-Schroer for insightful discussions concerning MSM methodology. 
J.F.R. and T.B. are also thankful to the organizers and participants of the 2015 Winter School on Markov State Models and Molecular and Chemical Kinetics conference.
Funding from the SFB-TRR146 grant of the German Research Foundation (DFG) is gratefully acknowledged.

%%%%%%%%%%%%%%%%%%%%%%%%%%%%%%
\section*{References} %\vspace{-1cm} %\renewcommand{\refname}{Full
                      %Bibliography} %\bibliographystyle{plain}
                      %%\bibliographystyle{apsrev4-1}
%\renewcommand*{\bibfont}{\footnotesize} %\renewcommand*{\citenumfont}[1]{#1}
%\renewcommand*{\bibnumfmt}[1]{[#1]}
\bibliography{references_PSU,references_MPIP}

\clearpage

\end{document}